\documentclass[multphys,vecphys]{svmult}
\usepackage{amssymb}
\usepackage{makeidx}     % allows index generation
\usepackage{graphicx}    % standard LaTeX graphics tool
\usepackage{multicol}    % used for the two-column index
\makeindex             % used for the subject index
\def\beq{\begin{equation}}
\def\eeq{\end{equation}}

\def\bea{\begin{eqnarray}}
\def\eea{\end{eqnarray}}
\def\ba{\begin{array}}                  %array
\def\ea{\end{array}}

\def\a{\alpha}

\def\d{\delta}
\def\D{\Delta}

\def\g{\gamma}

\def\o{\omega}

\begin{document}

\title*{Astrophysical Bounds on\\ Planck Suppressed Lorentz
  Violation
  %\thanks{To be published in {\em Quantum Gravity
   %   Phenomenology, Proceedings of the 40th winter school
    %  in theoretical physics}, eds. Kowalski-Glikman,
    %Amelino-Camelia.  (Springer-Verlag, 2004)}
    }
\author{Ted Jacobson\inst{1},
Stefano Liberati\inst{2}\and David Mattingly\inst{3}}
% Please full name, not just initials!
\institute{Institut d'Astrophysique de Paris,
%98bis bd Arago, 75014 Paris,
France, and \\
Department of Physics, University of Maryland,
%College Park, MD 20742
USA \texttt{jacobson@umd.edu} \and SISSA
%, Via Beirut 2-4, 34014 Trieste, Italy
and INFN, Trieste, Italy \texttt{liberati@sissa.it}
\and Department of Physics, University of California at Davis, USA
\texttt{mattingly@physics.ucdavis.edu}}%
\maketitle

This article reviews many of the observational constraints on Lorentz
symmetry violation (LV).  We first describe the GZK cutoff and other
phenomena that are sensitive to LV.  After a brief historical sketch of
research on LV, we discuss the effective field theory description of LV
and related questions of principle, technical results, and observational
constraints.  We focus on constraints from high energy astrophysics
on mass dimension five operators that contribute to LV 
electron and photon dispersion relations at order $E/M_{\rm Planck}$.
We also briefly discuss constraints on renormalizable 
operators, and review the current and future contraints on LV at order 
$(E/M_{\rm Planck})^2$.  

\section{Windows on quantum gravity?}
\label{sec:intro}

In most fields of physics it goes without saying that observation
and prediction play a central role, but unfortunately quantum
gravity (QG) has so far not fit that mold.  Many intriguing and
ingenious ideas have been explored, but it seems safe to say that
without both observing phenomena that depend on QG, and extracting
reliable predictions from candidate theories that can be compared
with observations, the goal of a theory capable of incorporating
quantum mechanics and general relativity will remain unattainable.

Besides the classical limit, there is one observed phenomenon for
which quantum gravity makes a prediction that has received
encouraging support: the
spectrum of primordial
cosmological perturbations. The quantized longitudinal linearized
gravitational mode, albeit slave to the inflaton and not a
dynamically independent degree of freedom, plays an essential role
in this story~\cite{Mukhanov:1990me}.

What other types of phenomena might be characteristic of a quantum
gravity theory?  Motivated by tentative theories, partial
calculations, intimations of symmetry violation, hunches,
philosophy, etc, some of the proposed ideas are: loss of quantum
coherence or state collapse, QG imprint on initial cosmological
perturbations, scalar moduli or other new fields, extra dimensions
and low-scale QG, deviations from Newton's law, black holes
produced in colliders, violation of global internal symmetries,
and violation of spacetime symmetries. It is this last item, more
specifically the possibility of Lorentz violation (LV), that is
the focus of these lecture notes.

{From} the observational point of view, developments are encouraging
a new look at the possibility of LV.  Increased detector size,
space-borne instruments, technological improvement, and technique
refinement are permitting observations to probe higher energies,
weaker interactions, lower fluxes, lower temperatures, shorter
time resolution, and longer distances.  It comes as a welcome
surprise that the day of true quantum gravity observations may not
be so far off~\cite{Dawn}.

%%%%%%%%%%%%%%%%%%%%%%%%%%%%%%%%%%%%
\section{Lorentz violation}
\label{sec:LV?}
%%%%%%%%%%%%%%%%%%%%%%%%%%%%%%%%%%%%

Lorentz symmetry is linked to a scale-free nature of spacetime:
unbounded boosts expose ultra-short distances, and yet nothing
changes.  However, suggestions for Lorentz violation have come
from: the need to cut off UV divergences of quantum field theory
and of black hole entropy, tentative calculations in various QG
scenarios (e.g. semiclassical spin-network calculations in Loop
QG, string theory tensor VEVs, non-commutative geometry, some
brane-world backgrounds), and the possibly missing GZK
cutoff~\cite{GZK} on ultra-high energy (UHE) cosmic rays.

The GZK question has generated a lot of interest, and is
currently the only observational phenomenon thought to
indicate a possible violation of Lorentz symmetry.
As an invitation
to the subject, we discuss it in this section, before embarking on
the rest of the lectures.
We also give a list of possible LV phenomena, and a brief
historical overview of the subject.

\subsection{The GZK cut-off}
\index{GZK cutoff} In collisions of ultra high energy protons with
cosmic microwave background (CMB) photons there can be sufficient
energy in the center of mass frame to create a pion, leading to
the the reaction
\begin{equation}
p+\g_{\rm CMB}\rightarrow p+\pi^0.
\end{equation}
%
% new:
The threshold occurs when the invariant magnitude of the total
four momentum is the sum of the proton and pion mass, since at
threshold these particles are both at rest in the zero-momentum
frame. That is, it occurs when $(p_p+p_\g)^2=(m_p+m_\pi)^2$, or
$p_p\cdot p_\g=m_pm_\pi+\frac12m_\pi^2$, where $p_{p,\g}$ are the
proton and photon 4-momenta, and $m_{p,\pi}$ are the proton and
pion mass. Since $E_p\gg m_p$, and $m_\pi\ll 2m_p$, this yields
the proton energy threshold
\beq E_{GZK}\simeq \frac{m_p m_\pi}{2E_\g} \simeq 3\times 10^{20}
{\rm eV}\times\left(\frac{2.7{\rm K}}{E_\g}\right) \label{EGZK}
\eeq
To get a definite number we have put $E_\g$ equal to the energy of
a photon at the CMB temperature, 2.7{\rm K}, but of course the CMB
contains photons of higher energy,

This process degrades the initial proton energy with an
attenuation length of about 50 Mpc.  Since plausible astrophysical
sources for UHE particles are located at distances larger than 50
Mpc, one expects a cutoff in the cosmic ray proton energy
spectrum, which occurs at around $5\times10^{19}$ eV,
depending on the distribution of sources~\cite{Stecker:2003wm}.

\index{GZK cutoff!possible absence of} One of the experiments
measuring the UHE cosmic ray spectrum, the AGASA experiment, has
not seen the cutoff. An analysis~\cite{DeMarco:2003ig} from
January 2003 concluded that the cutoff was absent at the 2.5 sigma
level, while another experiment, HiRes, is consistent with the
cutoff but at a lower confidence level. (For a brief review of the
data see~\cite{Stecker:2003wm}.) The question should be answered
in the near future by the AUGER observatory, a combined array of
1600 water \v{C}erenkov detectors and 24 telescopic air
flouresence detectors under construction on the Argentine
pampas~\cite{Auger}. The new observatory will see an event rate
one hundred times higher, with better systematics.

Many ideas have been put forward to explain the possible absence
of the GZK cutoff~\cite{Stecker:2003wm}. For example the cosmic
rays might originate closer, in some unexpected way, by
astrophysical acceleration or by decay of ultra-heavy exotic
particles, or they may be produced by collisions with ultra high
energy cosmic neutrinos. Virtually all of these explanations have
problems.

\index{GZK cutoff!and Lorentz violation}In this context, it is
intriguing to consider that with even a tiny amount of Lorentz
violation the energy threshold for the GZK reaction could be
affected. According to equation~(\ref{EGZK}) the Lorentz invariant
threshold is proportional to the proton mass. Thus any LV term
added to the proton dispersion relation $E^2={\bf p}^2 + m^2$ will
modify the threshold if it is comparable to or greater than
$m_p^2$ at around the energy $E_{GZK}$. Modifying the proton and
pion dispersion relations, the threshold can be lowered, raised,
or removed entirely, or even an upper threshold where the reaction
cuts off could be introduced (see e.g.~\cite{Jacobson:2002hd} and
references therein).

For example, the LV term considered by Coleman and
Glashow~\cite{CG-GZK} was of the form $\eta {\bf p}^2$, assumed
given in a reference frame close to that of the earth, which is
natural since we are close to being at rest in the universal rest
frame. This would affect the GZK threshold as long as
$\eta>(m_p/E_{GZK})^2\sim 10^{-22}$. Even LV suppressed by two
powers of the Planck mass $M$ would affect the threshold: a term
of the form ${\bf p}^4/M^2$ is comparable to $m_p^2$ when the
proton energy is $(m_p M)^{1/2}\simeq3\times10^{18}$ eV, which is
two orders of magnitude below the highest energy cosmic rays. Thus
a missing GZK cutoff could be explained by Planck
double-suppressed LV. Conversely, observational confirmation of
the GZK cutoff can severely constrain such LV.

\subsection{Possible LV phenomena}

Trans-GZK cosmic rays are not the only window of opportunity we
have to detect or constrain Lorentz violation induced by QG
effects.  In fact, many phenomena accessible to current
observations/experiments are sensitive to possible violations of
Lorentz invariance. A partial list is
\begin{itemize}
\item sidereal variation of LV couplings as the lab moves
  with respect to a preferred frame or directions
\item long baseline dispersion and vacuum birefringence
  (e.g.~of signals from gamma ray bursts, active galactic
  nuclei, pulsars, galaxies)
\item new reaction thresholds (e.g.~photon decay, vacuum
\v{C}erenkov effect) \item shifted thresholds (e.g.~photon
annihilation from
  blazars, GZK reaction)
\item maximum velocity (e.g.~synchrotron peak from supernova
remnants) \item dynamical effects of LV background fields (e.g.
  gravitational coupling and additional wave modes)
\end{itemize}

\subsection{A brief history of some LV research}

\index{Lorentz violation!history of research}We conclude this
section with a brief historical overview mentioning some of the
more influential papers, but by no means complete.

Suggestions of possible LV in particle physics go back at least to
the 1960's, when a number of authors wrote on that
idea~\cite{Dirac}\footnote{
Remarkably, already in 1972
Kirzhnits and Chechin~\cite{Dirac} explored the
  possibility that an apparent missing cutoff in the UHE
  cosmic ray spectrum could be explained by something
  that looks very similar to the recently proposed
  ``doubly special relativity''~\cite{DSR}.}.  The
possibility of LV in a metric theory of gravity was explored
beginning at least as early as the 1970's~\cite{LVmetr}. Such
theoretical ideas were pursued in the '70's and '80's notably by
Nielsen and several other authors on the particle theory
side~\cite{7080theory}, and by Gasperini~\cite{Gasp} on the
gravity side.  A number of observational limits were obtained
during this period~\cite{HauganAndWill}.

Towards the end of the 80's Kostelecky and Samuel~\cite{KS}
presented evidence for possible spontaneous LV in string theory,
and motivated by this explored LV effects in gravitation.  The
role of Lorentz invariance in the ``trans-Planckian puzzle" of
black hole redshifts and the Hawking effect was emphasized in the
early 90's~\cite{TedUltra}. This led to study of the Hawking
effect for quantum fields with LV dispersion relations commenced
by Unruh~\cite{Unruh} and followed up by others. Early in the
third millenium this line of research led to work on the related
question of the possible imprint of trans-Planckian frequencies on
the primordial fluctuation spectrum~\cite{BM}.  Meanwhile the
consequences of LV for particle physics were being explored using
LV dispersion relations e.g. by Gonzalez-Mestres~\cite{GM}.

Four developments in the late nineties seem to have stimulated a
surge of interest in LV research. One was
a systematic extension of the standard model of particle physics
incorporating all possible LV in the renormalizable sector,
developed by Colladay and Kosteleck\'{y}~\cite{CK}.  That provided
a framework for computing the observable consequences for any
experiment and led to much experimental work setting limits on the
LV parameters in the lagrangian~\cite{AKbook}. On the
observational side, the AGASA experiment reported events beyond
the GZK cutoff~\cite{Takeda:1998ps}. Coleman and Glashow then
suggested the possibility that LV was the culprit in the possibly
missing GZK cutoff~\cite{CG-GZK}, and explored many other high
energy consequences of renormalizable, isotropic LV leading to
different limiting speeds for different particles~\cite{CGlong}.
In the fourth development,
it was pointed out by Amelino-Camelia et al~\cite{GAC-Nat} that
the sharp high energy signals of gamma ray bursts could reveal LV
photon dispersion suppressed by one power of energy over the mass
$M\sim 10^{-3}M_{\rm P}$, tantalizingly close to the Planck mass.

Together with the improvements in observational reach mentioned
earlier, these developments attracted the attention of a large
number of researchers to the subject. Shortly after
Ref.~\cite{GAC-Nat} appeared, Gambini and Pullin~\cite{GP} argued
that semiclassical loop quantum gravity suggests just such LV.
Some later work supported this notion, but the issue continues to
be debated~\cite{KP,Alfaro:2004ur}. In any case, the dynamical
aspect of the theory is not under enough control at this time 
to make any definitive statements concerning LV.

A very strong constraint on photon birefringence was obtained by
Gleiser and Kozameh~\cite{GK} using UV light from distant
galaxies. If the recent report\cite{CB03} of polarized gamma rays
from a GRB
turns out to be correct despite the concerns of Refs.~\cite{RF03},
this constraint will be
further strengthened dramatically~\cite{JLMS,Mitro}.  Further
stimulus came from the suggestion~\cite{PM} that an LV threshold
shift might explain the apparent under-absorption on the cosmic IR
background of TeV gamma rays from the blazar Mkn501, however it is
now believed by many that this anomaly goes away when a corrected
IR background is used~\cite{Kono}.

The extension of the effective field theory framework to include
LV dimension 5 operators was introduced by Myers and
Pospelov~\cite{MP}, and used to strengthen prior constraints. Also
this framework was used to deduce a very strong
constraint~\cite{Crab} on the possibility of a maximum electron
speed less than the speed of light from observations of
synchrotron radiation from the Crab Nebula.

\section{Theoretical framework for LV}

\index{Lorentz violation!theoretical framework}
Various different theoretical approaches to LV have been taken to
pursue the ideas summarized above.  Some researchers restrict
attention to LV described in the framework of effective field
theory (EFT), while others allow for effects not describable in
this way, such as those that might be due to stochastic
fluctuations of a ``space-time foam''. Some restrict to
rotationally invariant LV, while others consider also rotational
symmetry breaking. Both true LV as well as ``deformed" Lorentz
symmetry (in the context of so-called ``doubly special
relativity"\cite{DSR}) have been pursued. Another difference in
approaches is whether one allows for distinct LV parameters for
different particle types, or proposes a more universal form of LV.

\index{Lorentz violation!effective field theory|(}The rest of this article will
focus on just one of these approaches, namely LV describable by
standard EFT, assuming rotational invariance, and allowing
distinct LV parameters for different particles.  In exploring the
possible phenomenology of new physics, it seems useful to retain
enough standard physics so that clear predictions can be made, and
so that the possibilities are narrow enough to be meaningfully
constrained.

This approach is not universally favored. For example a sharp
critique appears in~\cite{GAC-crit}. Therefore we think it is
important to spell out the motivation for the choices we have
made. First, while of course it may be that EFT is not adequate
for describing the leading quantum gravity phenomenology effects,
it has proven itself very effective and flexible in the past. It
produces local energy and momentum conservation laws, and seems to
require for its validity just locality and local spacetime
translation invariance above some length scale. It describes the
standard model and general relativity (which are presumably not
fundamental theories), a myriad of condensed matter systems at
appropriate length and energy scales, and even string theory (as
perhaps most impressively verified in the calculations of black
hole entropy and Hawking radiation rates).  It is true that, e.g.,
non-commutative geometry (NCG) seems to lead to EFT with
problematic IR/UV mixing, however this more likely indicates a
physically unacceptable feature of such NCG rather than a physical
limitation of EFT.

\index{Lorentz violation!rotational invariance}
The assumption of rotational invariance is motivated by the idea
that LV may arise in QG from the presence of a short distance
cutoff. This suggests a breaking of boost invariance, with a
preferred rest frame, but not necessarily rotational invariance.
Since a constraint on pure boost violation is, barring a
conspiracy, also a constraint on boost plus rotation violation, it
is sensible to simplify with the assumption of rotation invariance
at this stage. The preferred frame is assumed to coincide with the
rest frame of the CMB.

\index{Lorentz violation!equivalence principle}
Finally
why do we choose to complicate matters by allowing for different
LV parameters for different particles? First, EFT for first order
Planck suppressed LV (see section~\ref{subsec:EFTLV}) requires
this for different polarizations or spin states, so it is
unavoidable in that sense. Second, we see no reason {\it a priori}
to expect these parameters to coincide.  The term ``equivalence
principle'' has been used to motivate the equality of the
parameters. However, in the presence of LV dispersion relations,
free particles with different masses travel on different
trajectories even if they have the same LV
parameters~\cite{Fischbach:wq,Jacobson:2002hd}.  Moreover,
different particles would presumably interact differently with the
spacetime microstructure since they interact differently with
themselves and with each other. An example of this occurs in the
braneworld model discussed in Ref.~\cite{Burgess}, and an extreme
version occurs in the proposal of Ref.~\cite{emn} in which only
certain particles feel the spacetime foam effects. (Note however
that in this proposal the LV parameters fluctuate even for a given
kind of particle, so EFT would not be a valid description.)

\subsection{Deformed dispersion relations}
\label{subsec:disp}

\index{Lorentz violation!dispersion relations}
A simple approach to a
phenomenological description of LV is via deformed dispersion
relations. If rotation invariance and integer powers of momentum
are assumed in the expansion of $E^2({\bf p})$, the dispersion
relation for a given particle type can be written as
\begin{equation}
E^2=p^2 + m^2 + \Delta(p), \label{gen-disprel}
\end{equation}
where $p$ is hereafter the magnitude of the three-momentum, and
\beq \Delta(p)= \tilde{\eta}_1 p^1 + \tilde{\eta}_2 p^2 +
\tilde{\eta}_3 p^3 + \tilde{\eta}_4 p^4 +\cdots \label{disprel1}
\eeq
Since they are not Lorentz invariant, it is necessary to specify
the frame in which these relations are given, namely the CMB
frame.

Let us introduce two mass scales, $M=10^{19}\, {\rm
  GeV}\approx M_{\rm Planck}$, the putative scale of
quantum gravity, and $\mu$, a particle physics mass scale.  To
keep mass dimensions explicit we factor out possibly appropriate
powers of these scales, defining the dimensionful $\tilde{\eta}$'s in
terms of corresponding dimensionless parameters. It might seem
natural that the $p^n$ term with $n\ge3$ be suppressed by
$1/M^{n-2}$, and indeed this has been assumed in most work. But
following this pattern one would expect the $n=2$ term to be
unsuppressed and the $n=1$ term to be even more important. Since
any LV at low energies must be small, such a pattern is untenable.
Thus either there is a symmetry or some other mechanism protecting
the lower dimension oprators from large LV, or the suppression of
the higher dimension operators is greater than $1/M^{n-2}$.  This
is an important issue to which we return in
section~\ref{naturalness}.

For the moment we simply follow the observational lead and insert
at least one inverse power of $M$ in each term, viz.
\beq \tilde{\eta}_1=\eta_1 \frac{\mu^2}{M},\qquad
\tilde{\eta}_2=\eta_2 \frac{\mu}{M},\qquad \tilde{\eta}_3=\eta_3
\frac{1}{M},\qquad \tilde{\eta}_4=\eta_4 \frac{1}{M^2}.
\label{disprel2} \eeq
%
%t30
In characterizing the strength of a constraint we refer to the
$\eta_n$ without the tilde, so we are comparing to what might be
expected from Planck-suppressed LV. We allow the LV parameters
$\eta_i$ to depend on the particle type, and indeed it turns out
that they {\it
  must} sometimes be different but related in certain
ways for photon polarization states, and for particle and
antiparticle states, if the framework of effective field theory is
adopted. In an even more general setting, Lehnert~\cite{Leh}
studied theoretical constraints on this type of LV and deduced the
necessity of some of these parameter relations.

The deformed dispersion relations are introduced for elementary
particles only; those for macroscopic objects are then inferred by
addition. For example, if $N$ particles with momentum $\bf p$ and
mass $m$ are combined, the total energy, momentum and mass are
$E_{\rm tot}=NE(p)$, ${\bf p}_{\rm tot}=N{\bf p}$, and $m_{\rm
tot}=Nm$, so that $E_{\rm tot}^{2}=p_{\rm tot}^{2} + m_{\rm
tot}^{2}+ N^2\D(p)$.  Although the Lorentz violating term can be
large in some fixed units, its ratio with the mass and momentum
squared terms in the dispersion relation is the same as for the
individual particles. Hence, there is no observational conflict
with standard dispersion relations for macroscopic objects.

This general framework allows for superluminal propagation, and
spacelike 4-momentum relative to a fixed background metric.  It
has been argued~\cite{KL} that this leads to problems with
causality and stability. In the setting of a LV theory with a
single preferred frame, however, we do not share this opinion. As
long as the physics is guaranteed to be causal and the states all
have positive energy in the preferred frame, we cannot see any
room for such problems to arise.

\subsection{Effective field theory and LV} \label{subsec:EFTLV}
\index{Lorentz violation!effective field theory!renormalizable operators}The
standard model extension (SME) of Colladay and
Kosteleck\'{y}~\cite{CK} consists of the standard model of
particle physics plus all Lorentz violating renormalizable
operators (i.e. of mass dimension $\le4$) that can be written
without changing the field content or violating the gauge
symmetry. For illustration, the leading order terms in the QED
sector are the dimension three terms
\beq -b_a\bar{\psi}\gamma_5 \gamma^a \psi
-\frac{1}{2}H_{ab}\bar{\psi}\sigma^{ab}\psi \eeq
and the dimension four terms
\beq -\frac{1}{4}k^{abcd}F_{ab}F_{cd}
+\frac{i}{2}\bar{\psi}(c_{ab}+ d_{ab}\gamma_5)\gamma^a
\stackrel{\leftrightarrow}{D}\!\!{}^b\psi, \eeq
where the dimension one coefficients $b_a$, $H_{ab}$ and
dimensionless $k^{abcd}$, $c_{ab}$, and $d_{ab}$ are constant
tensors characterizing the LV.  If we assume rotational invariance
then these must all be constructed from a given unit timelike
vector $u^a$ and the Minkowski metric $\eta_{ab}$, hence
$b_a\propto u_a$, $H_{ab}=0$, $k^{abcd}\propto u^{[a} \eta^{b][c}
u^{d]}$, $c_{ab}\propto u_au_b$, and $d_{ab}\propto u_au_b$.
Such LV is thus characterized by just four numbers.

\index{Lorentz violation!effective field theory!higher dimension operators}
The study
of Lorentz violating EFT in the higher mass dimension sector was
initiated by Myers and Pospelov~\cite{MP}.  They classified all LV
dimension five operators that can be added to the QED Lagrangian
and are quadratic in the same fields, rotation invariant, gauge
invariant, not reducible to a combination of lower and/or higher
dimension operators using the field equations, and contribute
$p^3$ terms to the dispersion relation. Just three operators
arise:
\beq
-\frac{\xi}{2M}u^mF_{ma}(u\cdot\partial)(u_n\tilde{F}^{na})+\frac{1}{M}
u^m\bar{\psi}\gamma_m(\zeta_1+\zeta_2\gamma_5)(u\cdot
\partial)^2\psi \label{dim5} \eeq
where $\tilde{F}$ denotes the dual of $F$, and $\xi,\zeta_{1,2}$
are dimensionless parameters. The sign of the $\xi$ term in
(\ref{dim5}) is opposite to that in~\cite{MP}, and is chosen
so that positive helicity photons have $+\xi$ for a dispersion
coefficient (see below). All of these terms violate CPT symmetry
as well as Lorentz invariance. Thus if CPT were
preserved, these LV operators would be forbidden.

\index{Lorentz violation!dispersion relations}
In the limit of high energy $E\gg m$, the photon and electron
dispersion relations following from QED with the above terms
are~\cite{MP}
\bea
\omega_{\pm}^2&=& k^2 \pm \frac{\xi}{M}k^3\\
E_{\pm}^2&=& p^2 + m^2  +\frac{2(\zeta_1\pm\zeta_2)}{M}p^3.
\label{QEDdisp} \eea
\index{Lorentz violation!dispersion relations!helicity dependence}
The photon subscripts $\pm$ refer to helicity, i.e. right and left
circular polarization,
which it turns out necessarily have opposite LV parameters. The
electron subscripts $\pm$ refer to the helicity, which can be
shown to be a good quantum number in the presence of these LV
terms~\cite{JLMS}.
\index{Lorentz violation!dispersion relations!positrons}
Moreover, if we write
$\eta_\pm=2(\zeta_1\pm\zeta_2)$ for the LV parameters of the two
electron helicities, those for positrons
are given~\cite{JLMS} by \beq \label{eq:ep} \eta^{\rm
positron}_\pm=-\eta^{\rm electron}_\mp. \eeq
If $\eta_1=0$, then the two helicities have opposite LV
paprameters, $\eta_+=-\eta_-$, so electron and positron have the
{\it same} LV parameters. If instead $\eta_2=0$, then the
$\eta_+=\eta_-$, so electron and positron have {\it opposite} LV
parameters.

\subsection{Naturalness of small LV at low energy?}
\label{naturalness}
\index{Lorentz violation!effective field theory!naturalness|(}
As discussed above in subsection
\ref{subsec:disp}, if LV operators of dimension $n>4$ are
suppressed, as we have imagined, by $1/M^{n-2}$, LV would feed
down to the lower dimension operators and be strong at low
energies~\cite{CGlong,MP,PerezSudarsky,Collins}, unless there is a
symmetry or some other mechanism that protects the lower dimension
operators from strong LV. What symmetry (other than Lorentz
invariance, of course!) could that possibly be?

\index{Lorentz violation!lattice field theory}
In the Euclidean context, a discrete subgroup of the Euclidean
rotation group suffices to protect the operators of dimension four
and less from violation of rotation symmetry. For
example~\cite{Hyper}, consider the ``kinetic'' term in the EFT for
a scalar field with hypercubic symmetry,
$M^{\mu\nu}\partial_\mu\phi\partial_\nu\phi$. The only tensor
$M^{\mu\nu}$ with hypercubic symmetry is proportional to the
Kronecker delta $\delta^{\mu\nu}$, so full rotational invariance
is an ``accidental'' symmetry of the kinetic operator.

If one tries to mimic this construction on a Minkowski lattice
admitting a discrete subgroup of the Lorentz group, one faces the
problem that each point has an infinite number of neighbors
related by the Lorentz boosts. For the action to share the
discrete symmetry each point would have to appear in infinitely
many terms of the discrete action, presumably rendering the
equations of motion meaningless.

\index{Lorentz violation!rotational symmetry}
Another symmetry that could do the trick is three dimensional
rotational symmetry together with a symmetry between different
particle types. For example, rotational symmetry would imply that
the kinetic term for a scalar field takes the form
$(\partial_t\phi)^2-c^2(\nabla\phi)^2$, for some constant $c$.
Then, for multiple scalar fields, a symmetry relating the fields
would imply that the constant $c$ is the same for all, hence the
kinetic term would be Lorentz invariant with $c$ playing the role
of the speed of light.  Unfortunately this mechanism does not work
in nature, since there is no symmetry relating all the physical
fields.

\index{Lorentz violation!supersymmetry} Perhaps under some
conditions a partial symmetry could be adequate, e.g. grand
unified gauge and/or super symmetry. In fact, a recent analysis of
Nibbelink and Pospelov~\cite{Nibbelink:2004za} presents evidence
that supersymmetry (SUSY) together with gauge symmetry might
indeed play this role. SUSY here refers to the symmetry algebra
that is a kind of square root of the spacetime translation group.
The nature of this square root depends upon the Minkowski metric,
so is tied to the Lorentz group, but it does not require Lorentz
symmetry. It is shown in Ref.~\cite{Nibbelink:2004za}, using the
superfield formalism, that the SUSY preserving LV operators that
can be added to the SUSY Standard Model first appear at dimension
five. Moreover, these operators do not contribute $O(p^3)$ terms
to the particle dispersion relations. Of course SUSY is broken in
the real world, but the suppression in the SUSY theory may mean
that the low dimension LV terms allowed in the presence of soft
SUSY breaking are suppressed enough to be compatible with
observation. On the other hand, it might also mean that they are
so suppressed as to lie beyond the scope of observation.

At this stage we assume the existence of some realization of the
Lorentz symmetry breaking scheme upon which constraints are being
imposed. If none exists, then our parametrization (\ref{disprel2})
is misleading, since there should be more powers of $1/M$
suppressing the higher dimension terms. In that case, current
observational limits on those terms do not significantly constrain
the fundamental theory.
\index{Lorentz violation!effective field theory!naturalness|)}
\index{Lorentz violation!effective field theory|)}

\section{Reaction thresholds and LV}
\label{sec:thresholds}

Lorentz violation can have significant effects on energy thresholds
for particle reactions. Such effects could be signatures
of LV, and can be used to put constraints on LV.
In the presence of LV, standard properties of LI
threshold configurations (e.g. angles and momentum
distributions) may not be preserved. Hence a careful study of
properties of LV threshold configurations is needed
before signatures and constraints can be considered. In this
section we review some basic results concerning LV thresholds.

\index{Lorentz violation!thresholds!general properties|(}
Threshold configurations
and new phenomena in the presence of LV dispersion
relations were systematically investigated in~\cite{CGlong,
Mattingly:2002ba} (see also references therein). We give here a
brief summary of the results. We shall consider reactions with
two initial and two final particles (results for reactions with
only one incoming or outgoing particle can be obtained as special
cases).  Following our previous choice of EFT
we allow each particle to have an independent dispersion relation
of the form (\ref{gen-disprel}) with $E(p)$ a  monotonically
increasing non-negative function of the magnitude $p$
of the 3-momentum $\bf p$.  While the assumption of
monotonicity could perhaps be violated at the Planck scale, it is
satisfied for any reasonable low energy expansion of a LV
dispersion relation. EFT further implies that energy and momentum
are additive for multiple particles, and conserved.

Consider a four-particle interaction where a target particle of
3-momentum ${\bf p}_2$ is hit by a particle of 3-momentum ${\bf p}_1$,
with an angle $\alpha$ between the two momenta, producing two
particles of momenta ${\bf p}_3$ and ${\bf p}_4$. We call $\beta$ the angle
between ${\bf p}_3$ and the total incoming 3-momentum
${\bf p}_{in}={\bf p}_1+{\bf p}_2$.
We define the notion of a threshold relative to a fixed value
of the magnitude of the target momentum $p_2$.
A lower or upper
threshold corresponds to a value of $p_1$
(or equivalently the energy $E_1$)
above which the reaction starts or stops
being allowed by energy and momentum conservation.

We now
introduce a graphical interpretation of the energy-momentum
conservation equation that allows the properties of thresholds
to be easily understood.
For given values of
$(p_1, p_2,\alpha,\beta, p_3)$, momentum conservation determines
$p_4$. Since $p_3$ and $p_4$ determine the final energies
$E_3$ and $E_4$, we can thus
define the final energy function
$E_f^{\alpha,\beta,p_3}(p_1)$. (Since $p_2$ is fixed
we drop it from the labelling.) Energy conservation
requires that $E_f$ be equal to $E_i(p_1)$, the
initial energy (again, we do not indicate the dependence
on the fixed momentum $p_2$).

\begin{figure}[htb]
\vbox{ \hfil \scalebox{0.60}{{\includegraphics{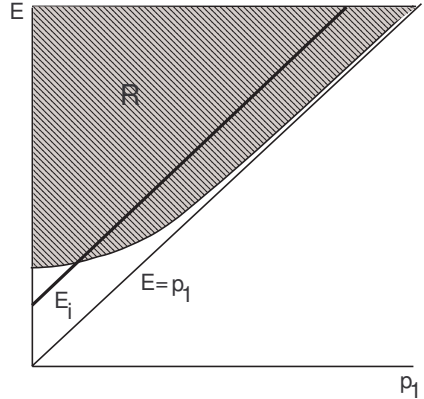}}} \hfil }
\bigskip
\caption{$\cal R$  is the region covered by all final energy curves
$E_f^{\alpha,\beta,p_3} (p_1)$ for some fixed $p_2$, with 
$p_4$ determined by momentum conservation. The curve
$E_i(p_1)$ is the initial energy for the same fixed $p_2$. Where
the latter curve lies inside  $\cal R$  there is a solution to the
energy and momentum conservation equations.} \label{fig:R}
\end{figure}

Now consider the region $\cal{R}$ in the $(E, p_1)$ plane covered
by plotting $E_f^{\alpha,\beta,p_3}(p_1)$  for all possible
configurations  $(\alpha,\beta, p_3)$. An example
is shown in Figure~\ref{fig:R}.
The region $\cal{R}$ is
bounded below by $E=0$ since the particle energies are
assumed non-negative, hence it has some bounding curve
$E_B(p_1)$. Similarly one can plot $E_i(p_1)$.
The reaction is allowed (i.e. there is a
solution to the energy and momentum conservation equations) when
this latter curve lies inside the region $\cal R$. A lower or upper
threshold occurs when $E_i(p_1)$ enters or leaves $\cal R$.

%=====================================================================
%=====================================================================

This graphical representation demonstrates that in any threshold
configuration (lower or upper) occuring at some $p_1$,
the parameters $(\alpha,\beta, p_3)$
are such that the final energy function $E_f^{\alpha,\beta,p_3}(p_1)$
is minimized. That is, the configuration always yields the
minimum final particle energy configuration conserving momentum
at fixed $p_1$ and $p_2$. From this fact, it is easy to
deduce two general properties of these configurations:
\index{Lorentz violation!thresholds!kinematic configurations}
\begin{enumerate}
\item All thresholds for processes with two outgoing
particles occur at parallel final momenta ($\beta=0$).
\item For a two-particle initial state the momenta are antiparallel at
threshold ($\alpha=\pi$).
\end{enumerate}
These properties are in agreement with the well known case of
Lorentz invariant kinematics. Nevertheless, LV thresholds can
exhibit new features not present in the Lorentz invariant theory,
in particular upper thresholds and asymmetric pair creation.

\index{Lorentz violation!thresholds!upper thresholds}
Figure~\ref{fig:R} clearly
shows that LV allows for a reaction to not only to start
at some lower threshold but also to
end at some upper threshold where the curve $E_i$ exits the
region $\cal R$. It can even happen that $E_i$ enters and exits
$\cal R$ more than once, in which case there are what
one might call ``local" lower and and upper thresholds.

\index{Lorentz violation!thresholds!asymmetric pair creation}
Another interesting novelty is the possibility to have a (lower or
upper) threshold for pair creation with an unequal partition of
the initial momentum $p_{in}$ into the two outgoing particles
(i.e. $p_3\neq p_4\neq p_{in}/2$). Equal partition of momentum is
a familiar result of Lorentz invariant physics, which follows from
the fact that the final particles are all at rest in the
zero-momentum frame at threshold. This has often been
(erroneously) presumed to hold as well in the presence of LV
dispersion relations.

\begin{figure}[htb]
\bigskip
\vbox{ \hfil \scalebox{0.60}{{\includegraphics{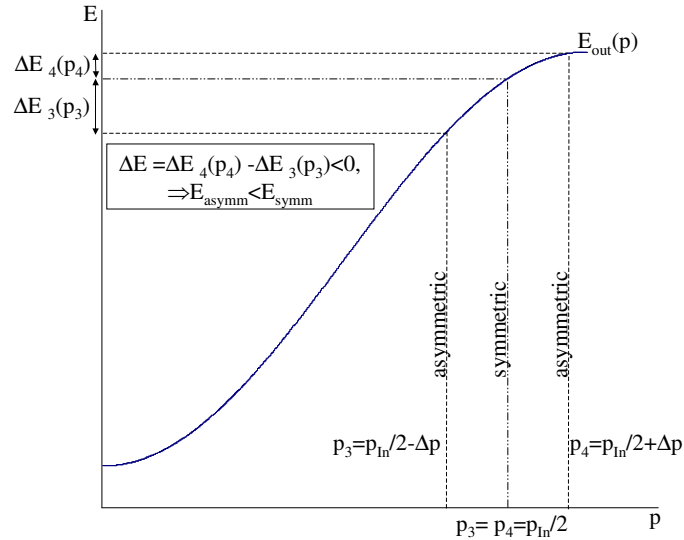}}} \hfil }
\bigskip
\caption{ Asymmetric pair production. The negative curvature of
the outgoing particle dispersion relation allows 
the energy of the outgoing pair to be reduced
by distributing the initial momentum $p_{in}$
un-equally between the two particles. } \label{fig:asymm}
\end{figure}

A reason for the occurrence of
asymmetric LV thresholds can be seen graphically,
as shown in Figure~\ref{fig:asymm}. Suppose the dispersion relation
for a massive outgoing particle $E_{out}({\bf p})$ has negative
curvature at $p=p_{in}/2$,
as might be the case for negative LV coefficients. Then
a small momentum-conserving displacement from a symmetric
configuration can lead to a net decrease in the final
state energy. According to the result established above,
the symmetric configuration cannot be the threshold one in such a case.
A lower $p_1$ could satisfy both
energy and momentum conservation with an asymmetric
final configuration.
%=====================================================================
%=====================================================================
A {\it sufficient} condition for the pair-creation threshold
configuration to be asymmetric is that the
final particle dispersion relation has negative curvature
at $p=p_{in}/2$. This condition is not necessary however, since
it could happen that the energy is locally but not globally
minimized by the symmetric configuration.

\section{Constraints}
\index{Lorentz violation constraints|(}
Observable effects of LV arise, among other
things, from 1)
sidereal variation of LV couplings due to motion of the laboratory
relative to the preferred frame, 2) dispersion and birefringence
of signals over long travel times, 3) anomalous reaction
thresholds.  We will often express the constraints in terms of the
dimensionless parameters $\eta_n$ introduced in (\ref{disprel2}).
An order unity value might be considered to be expected in Planck
suppressed LV.

\index{Lorentz violation!``amplifiers''}
The possibility of
interesting constraints in spite of Planck suppression arises in
different ways for the different types of observations. In the
laboratory experiments looking for sidereal variations, the
enormous number of atoms allow variations of a resonance frequency
to be measured extremely accurately. In the case of dispersion or
birefringence, the enormous propagation distances would allow a
tiny effect to accumulate. In the anomalous threshold case, the
creation of a particle with mass $m$ would be strongly affected by
a LV term when the momentum becomes large enough for this term to
be comparable to the mass term in the dispersion relation.

We briefly mention first constraints on the
renormalizable Standard Model Extension, then
focus on LV suppressed by one or two powers of
the ratio $E/M$.

\subsection{Constraints on renormalizable terms}
\index{Lorentz violation constraints!renormalizable terms}
For the $n=2$ term in (\ref{disprel1},\ref{disprel2}),
the absence of a strong threshold effect yields a constraint
$\eta_2 \lesssim (m/p)^2(M/\mu)$.  If we consider protons and put
$\mu=m=m_p\sim 1$ GeV, this gives an order unity constraint when
$p\sim \sqrt{mM} \sim 10^{19}$ eV. Thus the GZK threshold, if
confirmed, can give an order unity constraint, but multi-TeV
astrophysics yields much weaker constraints.  The strongest
laboratory constraints on dimension three and four operators come
from clock comparison experiments using noble gas
masers~\cite{Bear:2000cd}.  The constraints limit a combination of
the coefficients for dimension three and four operators for the
neutron to be below $10^{-31}$ GeV (the dimension four
coefficients are weighted by the neutron mass, yielding a
constraint in units of energy).
For more on such constraints see e.g.~\cite{AKbook,Bluhm:2003ne}.
Astrophysical limits on photon
vacuum birefringence give a bound on the coefficients of dimension
four operators of $10^{-32}$~\cite{KM}.

\subsection{Summary of constraints on LV in QED at $O(E/M)$}
\index{Lorentz violation constraints!QED at $O(E/M)$|(}

Since we do not assume universal LV coefficients, different
constraints cannot be combined unless they involve just the same
particle types.  To achieve the strongest combined constraints it
is thus preferable to focus on processes involving a small number
of particle types. It also helps if the particles are very common
and easy to observe.  This selects electron-photon physics, i.e.
QED, as a useful arena.

The current constraints on the three LV parameters at order
$E/M$---one in the photon dispersion relation and two in the
electron dispersion relation---will now be summarized.  These are
equivalent to the parameters in the dimension five operators
(\ref{dim5}) written down by Myers and Pospelov.

For $n=3$, a strong effect on energy thresholds involving only
electrons and photons can occur when the LV term $\eta p^3/M$ in
the electron or photon dispersion relation is comparable to or
greater than the electron mass term $m^2$. This happens when
\beq p\simeq 14\, {\rm TeV}\, \eta_3^{-1/3}. \label{1/p3} \eeq
We can thus obtain order unity and even much stronger constraints
from high energy astrophysics, where such energies are reached and
exceeded.

\begin{figure}[htb]
\vbox{ \hfil \scalebox{0.70}{{\includegraphics{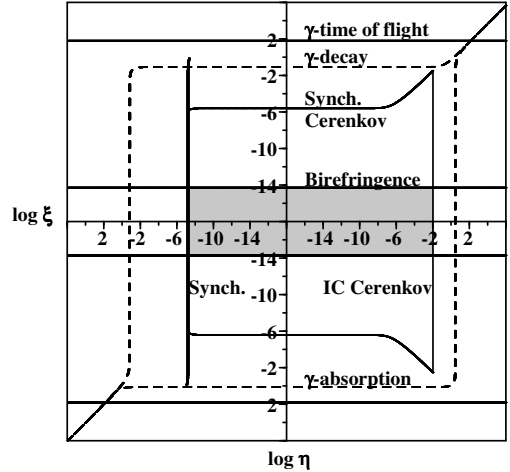}}} \hfil }
\bigskip
%\begin{figure}
%\center
%\includegraphics[height=8cm]{constraint1213}
%\picplace{5cm}{2cm} % Give the correct figure height and width in cm
%
\caption{Constraints on LV in QED at $O(E/M)$ (figure from
Ref.~\cite{JLMS}).}
\label{fig:1}       % Give a unique label
\end{figure}

In Fig.~\ref{fig:1} (from Ref.~\cite{JLMS}) constraints on the
photon ($\xi$) and electron ($\eta$) LV parameters are plotted on
a logarithmic scale to allow the vastly differing strengths to be
simultaneously displayed. For negative parameters, the negative of
the logarithm of the absolute value is plotted, and a region of
width $10^{-18}$ is excised around each axis. The synchrotron and
\v{C}erenkov constraints must both be satisfied by
%either $\eta_+$ or $\eta_-$.
at least one of the four quantities $\pm\eta_\pm$. The IC and
synchrotron \v{C}erenkov lines are truncated where they cross.
Prior photon decay and absorption constraints are shown in dashed
lines since they do not account for the EFT relations between the
LV parameters.

We now briefly review the physics and observations behind these
and other constraints.

%\paragraph
\subsubsection{Electron helicity dependence and ``helicity decay"}
\index{Lorentz violation constraints!QED at $O(E/M)$!helicity dependence} The
constraint $|\eta_+-\eta_-|<4$ on the difference between the
positive and negative electron helicity parameters was deduced by
Myers and Pospelov~\cite{MP} using a previous spin-polarized
torsion pendulum experiment~\cite{Heckel} that looked for diurnal
changes in resonance frequency.  They also determined a
numerically stronger constraint using nuclear spins, however this
involves four different LV parameters, one for the photon, one for
the up-down quark doublet, and one each for the right handed up
and down quark singlets. It also requires a model of nuclear
structure.

\index{Lorentz violation constraints!QED at $O(E/M)$!helicity decay}
\index{helicity decay}It is possible that an interesting
constraint could be obtained from the process of ``helicity
decay"\cite{JLMS}. If $\eta_+$ and $ \eta_-$ are unequal, say
$\eta_+>\eta_-$, then a positive helicity electron can decay into
a negative helicity electron and a photon, even when the LV
parameters do not permit the vacuum \v{C}erenkov effect. In this
process, the large $R$ or small ($O(m/E)$) $L$ component of a
positive helicity electron is coupled to the small $R$ or large
$L$ component of a negative helicity electron respectively.  Such
``helicity decay" has no threshold energy, so whether this process
can be used to set a constraint is solely a matter of the decay
rate.  It can be shown (assuming $|\xi|\lesssim10^{-3}$) that for
electrons of energy less than the transition energy
$(m^2M/(\eta_+-\eta_-))^{1/3}$, the lifetime of an electron
susceptible to helicity decay is greater than $4 \pi
M/(\eta_+-\eta_-)e^2 m^2$.  At the limit of the best current bound
$|\eta_+-\eta_-|<4$, the transition energy is approximately 10 TeV
and the lifetime for electrons below this energy is greater than
$10^4$ seconds.  This is long enough to preclude any terrestrial
experiments from seeing the effect. The lifetime above the
transition energy is instead bounded below by $E/e^2 m^2$, which
is $10^{-11}$ seconds for energies just above 10 TeV. The lifetime
might therefore be short enough to provide new constraints.
Such a constraint might come from the Crab Nebula, as explained
below.

%\paragraph
\subsubsection{Vacuum bifrefringence}
\index{Lorentz violation constraints!QED at $O(E/M)$!vacuum birefringence}
\index{vacuum birefringence}The birefringence constraint arises from the
fact that the LV parameters for left and right circular polarized
photons are opposite (\ref{QEDdisp}).  The phase velocity thus
depends on both the wavevector and the helicity. Linear
polarization is therefore rotated through an energy dependent
angle as a signal propagates, which depolarizes any initially
linearly polarized signal.  Hence the observation of linearly
polarized radiation coming from far away can constrain the
magnitude of the LV parameter.

In more detail, with the dispersion relation (\ref{QEDdisp}) the
direction of linear polarization is rotated through the angle
\beq \theta(t)=\left[\omega_+(k)-\omega_-(k)\right]t/2=\xi k^2
t/2M \label{rotation} \eeq
for a plane wave with wave-vector $k$ over a propagation time $t$.
The difference in rotation angles for wave-vectors $k_1$ and $k_2$
is thus
\beq
\Delta\theta=\xi (k_2^2-k_1^2) d/2M,
  \label{diffrotation}
\eeq
where we have replaced the time $t$ by the distance $d$ from the
source to the detector (divided by the speed of light). Note that
the effect is {\it quadratic} in the photon energy, and
proportional to the distance traveled.

This effect has been used to constrain LV in the dimension three
(Chern-Simons)~\cite{CFJ}, four~\cite{KM} and
five~\cite{GK,JLMS,Mitro} terms. The constraint shown in the
figure derives from the recent report~\cite{CB03} of a high degree
of polarization of MeV photons from GRB021206. The data analysis
has been questioned~\cite{RF03},
%and defended~\cite{CB03b},
so we shall have to wait and see if it is confirmed.
The next best
constraint on the dimension five term was deduced by Gleiser and
Kozameh~\cite{GK} using UV light from distant galaxies. While ten
orders of magnitude weaker, it is still very strong,
$|\xi|\lesssim 2\times10^{-4}$.

%\paragraph
\subsubsection{Photon time of flight}
\index{Lorentz violation constraints!QED at $O(E/M)$!photon time of flight}
Photon time of flight constraints~\cite{Schaefer} limit
differences in the arrival time at Earth for photons originating
in a distant event~\cite{Pavlopoulos,GAC-Nat}. Time of flight can
vary with energy since the LV term in the group velocity is $\xi
k/M$. The arrival time difference for wave-vectors $k_1$ and $k_2$
is thus
\beq
\Delta t=\xi (k_2-k_1) d/M,
  \label{timediff}
\eeq
which is proportional to the energy difference and the distance
travelled. Using the EFT result (\ref{QEDdisp}), the velocity
difference of the two polarizations at a given energy is
$2|\xi|k/M$, at least twice as large as the one arising from
energy differences. However, the time of flight constraint remains
many orders of magnitude weaker than the birefringence one from
polarization rotation. In Fig.~\ref{fig:1} we use the EFT
improvement of the constraint of Biller et al.~\cite{Schaefer}
(this is the best constraint to date for which a reliable
distance is known),
which yields $|\xi|<63$.

%\paragraph
\subsubsection{Vacuum \v{C}erenkov effect, inverse Compton electrons}
\index{Lorentz violation constraints!QED at $O(E/M)$!vacuum \v{C}erenkov effect}
\index{vacuum \v{C}erenkov effect} In the presence of LV the
process of vacuum \v{C}erenkov radiation $e\rightarrow e\gamma$
can occur. For example, if the photon dispersion is unmodified and
the electron parameter $\eta$ (for one helicity) is positive, then
the electron group velocity $v_g=1-(m^2/2p^2)+(\eta p/M)+\cdots$
exceeds the speed of light when
\beq p_{\rm th} =(m^2M/2\eta)^{1/3}\simeq 11\, {\rm
TeV}\, \eta^{-1/3}. \label{pcerenkov} \eeq
This turns out to be the threshold energy for the vacuum
\v{C}erenkov process with emission of a zero energy photon, which
we call the soft \v{C}erenkov threshold. \index{Lorentz
violation!thresholds!hard vs. soft \v{C}erenkov} There is also the
possibility of a hard \v{C}erenkov
threshold\cite{Jacobson:2002hd,KonMaj}. For example, if the
electron dispersion is unmodified and the photon parameter $\xi$
is negative then at sufficiently high electron energy the emission
of an energetic positive helicity photon is possible. This hard
\v{C}erenkov threshold occurs at $p_{\rm th}=(-4m^2M/\xi)^{1/3}$,
and the emitted photon carries away half the incoming electron
momentum. It turns out that the threshold is soft when both
$\eta>0$ and $\xi\ge -3\eta$, while it is hard when both $\xi
<-3\eta$ and $\xi<\eta$. The hard threshold in the general case is
given by $p_{\rm th}=(-4m^2M(\xi+\eta)/(\xi-\eta)^2)^{1/3}$, and
the photon carries away a fraction $(\xi-\eta)/2(\xi+\eta)$ of the
incoming momentum. In the general case at threshold, neither the
incoming nor outgoing electron group velocity is equal to the
photon group velocity, so the hard \v{C}erenkov effect cannot
simply be interpreted as being due to faster than light motion of
a charged particle.

\index{Lorentz violation constraints!QED at $O(E/M)$!vacuum
\v{C}erenkov and inverse Compton electrons}
\index{inverse Compton scattering} The inverse Compton (IC)
\v{C}erenkov constraint uses the electrons of energy up to 50 TeV
inferred via the observation of 50 TeV gamma rays from the Crab
nebula which are explained by IC scattering.
(The implications of a possible high energy population
of positrons is discussed below.)
Since the vacuum
\v{C}erenkov rate is orders of magnitude higher than the IC
scattering rate, that process must not occur for these
electrons~\cite{CGlong,Jacobson:2002hd}. The absence of the soft
\v{C}erenkov threshold up to 50 TeV produces the vertical IC
\v{C}erenkov line in Fig. \ref{fig:1}.  One can see from
(\ref{pcerenkov}) that this yields a constraint on $\eta$ of order
$(11~{\rm TeV}/50~{\rm TeV})^3\sim 10^{-2}$.  It could be that
only one electron helicity produces the IC photons and the other
loses energy by vacuum \v{C}erenkov radiation. Hence we can infer
only that at least one of $\eta_+$ and $\eta_-$ satisfies the
bound.

We do not indicate the hard IC \v{C}erenkov threshold constraint
in Fig.~\ref{fig:1} since it is superceded by the hard synchrotron
\v{C}erenkov constraint discussed below.

%\paragraph
\subsubsection{Crab synchrotron emission}
\index{Lorentz violation constraints!QED at $O(E/M)$!Crab synchrotron|(} A
constraint complementary to the \v{C}erenkov one was derived
in~\cite{Crab} by making use of the very high energy
electrons that produce the highest frequency
synchrotron radiation in the Crab nebula. For negative values of
$\eta$ the electron has a maximal group velocity less than the
speed of light, hence there is a maximal synchrotron frequency
that can be produced regardless of the electron
energy~\cite{Crab}. In the Lorentz invariant case these electrons
must have an energy of at least 1500 TeV, which suggests that we
should be able to obtain a constraint many orders of magnitude
stronger than the IC \v{C}erenkov one. We now explain how this
indeed comes about.

\index{synchrotron radiation} Cycling electrons in a magnetic
field $B$ emit synchrotron radiation with a spectrum that sharply
cuts off at a frequency $\omega_c$ given by the formula
\begin{equation}
\omega_c=\frac{3}{2} eB\frac{\gamma^3(E)}{E}\, ,
\label{eq:opeaklv}
\end{equation}
where $\gamma(E)=(1-v^2(E)/c^2)^{-1/2}$. Here $v(E)$ is the
electron group velocity, and $c$ is the usual low energy speed of
light. (As shown in~\cite{Crab} the photon energy is low enough to
neglect any possible LV correction as long as $|\xi|\lesssim
10^{11}(-\eta)^{4/3}$.) The formula (\ref{eq:opeaklv}) is based on
the electron trajectory for a given energy in a given magnetic
field, the radiation produced by a given current, and the
relativistic relation between energy and velocity. As explained
in~\cite{Crab}, and also in some more detail
in~\cite{Jacobson:2003ty} (which was written in response to the
criticism of~\cite{GAC-crit}), only the last of these ingredients
is significantly affected by LV in the EFT framework we are
considering. Hence (\ref{eq:opeaklv}) holds in that framework.

In standard relativistic physics, $E=\g m$, so the energy
dependence in (\ref{eq:opeaklv}) is entirely through the factor
$\g^2$, which grows without bound as the energy grows.
In the LV
case, the maximum synchrotron frequency $\omega_c^{\rm max}$ is
obtained by maximizing $\o_c$ (\ref{eq:opeaklv}) with respect to
the electron energy, which amounts to maximizing $\gamma^3(E)/E$.
Using the difference of group velocities
\begin{equation}
c-v(E)\simeq \frac{m^2}{2E^2} -
\eta\, \frac{E}{M},
\label{eq:vdiff}
\end{equation}
we find that this maximization yields
\begin{equation}
\omega_c^{\rm max}=0.34 \, \frac{eB}{m}(-\eta m/M)^{-2/3}.
\label{eq:opeaklv2}
\end{equation}
This maximum frequency is attained at the energy $E_{\rm
max}=(-2m^2M/5\eta)^{1/3}=10\, (-\eta)^{-1/3}$ TeV.
This is higher
than the energy that produces the same cutoff frequency
in the Lorentz invariant case,
but only by a factor of order unity.

The rapid decay of synchrotron emission at frequencies larger than
$\omega_c$ implies that most of the flux at a given frequency in a
synchrotron spectrum is due to electrons for which $\omega_c$ is
above that frequency. Thus $\omega_c^{\rm max}$ must be greater
than the maximum observed synchrotron emission frequency $\o_{\rm
obs}$. This yields the constraint
\begin{equation}
\eta > - \frac{M}{m}\left(\frac{0.34\, eB}{m\o_{\rm
obs}}\right)^{3/2}. \label{eq:synchcon}
\end{equation}

The Crab synchrotron emission has been observed to extend at least
up to energies of about 100 MeV~\cite{AA96}, just before the
inverse Compton hump begins to contribute to the spectrum. The
magnetic field in the emission region has been estimated by
several methods which agree on a value between 0.15--0.6 mG (see
e.g. \cite{Hillas} and references therein.) Two of these methods,
radio synchrotron emission and equipartition of energy, are
insensitive to Planck suppressed Lorentz violation, hence we are
justified in adopting a value of this order for the purpose of
constraining Lorentz violation. We use the largest value 0.6 mG
for $B$, since it yields the weakest constraint.

Our prior work assumed the high energy Crab radiation was produced
purely by electrons, not positrons. We consider here first this
case. Then we infer that at least one of the two parameters
$\eta_\pm$ must be greater than $-7\times10^{-8}$. 
We cannot constrain both $\eta$ parameters in this way since it could be
that all the Crab synchrotron radiation is produced by electrons
of one helicity. 

\paragraph{Combined synchrotron \& IC \v{Cerenkov} constraint}
The $\eta$ satisfying the synchrotron constraint 
{\it must be the same $\eta$ as satisfies the IC
  \v{C}erenkov constraint discussed above}.
If the synchrotron $\eta$ did not satisfy the IC \v{C}erenkov
constraint, the energy of these synchrotron electrons would
necessarily be under 50 TeV, rather than over the Lorentz
invariant value of 1500 TeV.  The Crab spectrum is well accounted
for with a single population of electrons responsible for both the
synchrotron radiation and the IC $\g$-rays. If there were enough
extra electrons to produce the observed synchrotron flux with
thirty times less energy per electron, then the electrons of the
other helicity, would be equally numerous and would therefore
produce too many IC $\g$-rays~\cite{JLMS}. It is important that
the {\it same} $\eta$, i.e.\  either $\eta_+$ or $\eta_-$,
satisfies both the synchrotron and the IC \v{C}erenkov
constraints. Otherwise, both constraints could have been satisfied
by having one of these two parameters arbitrarily large and
negative, and the other arbitrarily large and
positive.\footnote{We thank G. Amelino-Camelia for  
focusing our attention on this point.}

\paragraph{Possible helicity dependence constraint}
\index{Lorentz violation constraints!QED at $O(E/M)$!helicity decay}
As alluded to above,
a constraint on helicity
dependence of the electron parameter $\eta$ 
might be possible using the Crab Nebula.
Suppose that $\eta_-$ is below the synchrotron constraint
(i.e. $\eta_-<-7\times10^{-8}$), so that $\eta_+$ must satisfy
both the synchrotron and \v{C}erenkov constraints
as explained above. Then
positive helicity electrons must have an energy of at
least 50 TeV to produce the observed synchrotron 
radiation. These must not decay to negative helicity
electrons (since those would be unable to produce the 
synchrotron emission). This would require that the transition energy
(discussed inthe helicity dependence section above) 
be greater than 50 TeV if the decay rate 
is fast enough. This would yield the constraint
$\eta_+-\eta_-<10^{-2}$.

\paragraph{Possible role of positrons}
If the population of high energy charges includes positrons as
well as electrons, as in some models\cite{alice}, then the above
constraint analysis must be modified. The reasoning discussed so
far implies only that at least one of the four parameters
$\pm\eta_\pm$ satisfies both the synchrotron and IC \v{C}erenkov
constraints, since the emitting charge could be either an electron
or a positron. In effect, this reduces to the statement that
one of $|\eta_\pm|$ satisfies the IC Cerenkov constraint.
We are currently investigating what constraints can
be inferred if the amount of radiation produced by each of the
four populations of charges is accounted for more quantitatively.
\index{Lorentz violation constraints!QED at $O(E/M)$!Crab synchrotron|)}

%\paragraph
\subsubsection{Vacuum \v{C}erenkov effect, synchrotron electrons}
\index{Lorentz violation constraints!QED at $O(E/M)$!vacuum
\v{C}erenkov effect and synchrotron electrons}
\index{vacuum \v{C}erenkov effect} The existence
of the synchrotron producing electrons can be exploited to extend
the vacuum \v{C}erenkov constraint. For a given $\eta$ satisfying
the synchrotron bound, some definite electron energy $E_{\rm
synch}(\eta)$ must be present to produce the observed synchrotron
radiation. (This is higher for negative $\eta$ and lower for
positive $\eta$ than the Lorentz invariant value~\cite{Crab}.)
Values of $|\xi|$ for which the vacuum \v{C}erenkov threshold is
lower than $E_{\rm synch}(\eta)$ for either photon helicity can
therefore be
excluded~\cite{JLMS}. %t30 added ref to us here.
This is always a hard photon threshold, since the soft photon
threshold occurs when the electron group velocity reaches the low
energy speed of light, whereas the velocity required to produce
any finite synchrotron frequency is smaller than this.

%\paragraph
\subsubsection{Photon decay}
\index{Lorentz violation constraints!QED at $O(E/M)$!photon decay}
\index{photon decay}In the presence of LV the process of photon decay
$\g\rightarrow e^+e^-$ can occur. For example, if the electron
dispersion is unmodified and the photon parameter $\xi$ is
positive, the positive helicity photon decays above the threshold
energy $k_{\rm th}=(4m^2M/\xi)^{1/3}$. If instead the photon
dispersion is unmodified and if electron and positron have the
same dispersion with $\eta<0$, then the threshold occurs at
$k_{\rm th}=(-8m^2M/\eta)^{1/3}$. The threshold for general $\xi$
and $\eta$ is found in Refs.~\cite{Jacobson:2002hd,KonMaj}.

Contrary to relativistic intuition, it turns out that when
$\eta<\xi<0$ the electron and positron are not created with the
same momentum. The reason ({\it cf.} section \ref{sec:thresholds})
is the electron and positron energy functions $E(p)$ have negative
curvature at the threshold value of $p$. If the two momenta were
equal, the energy of the final state at fixed momentum could be
lowered by making the momentum of one particle smaller and one
larger by an equal amount.

Previous work on observational constraints using photon decay and
photon absorption (to be discussed below) were carried out before
it was known how the dispersion depends on helicity and particle
vs. anti-particle. Since these constraints are in any case not
competitive now with others, we have not attempted to fully
account for these relations. Here we just make a few remarks.

The strongest limit on photon decay came from the highest energy
photons known to propagate, which at the moment are the 50 TeV
photons observed from the Crab
nebula~\cite{Jacobson:2002hd,KonMaj}. These photons must not decay
before reaching the earth, so we can rule out those LV parameters
that lead to a threshold below 50 TeV, provided the decay rate is
fast enough.

Since we do not know the polarization of the observed photons
however, we can only exclude regions where \textit{both} photon
polarizations decay. Recall that according to (\ref{QEDdisp})
positive and negative helicity photons have opposite parameters
$\pm\xi$. A positive helicity photon carries a spin angular
momentum of one along the direction of motion. At threshold, where
all momenta are aligned, the electron and positron must therefore
both have positive helicity. Likewise a left-handed photon decays
at threshold into a negative helicity pair. Consider first the
case $\eta_-=-\eta_+$ so that, according to (\ref{eq:ep}), the
electron and positron have the same dispersion parameter. Then the
outgoing pair both have parameter $\eta_+$ for a positive helicity
incoming photon and $-\eta_+$ for a negative helicity one. We can
then exclude those parameters for which {\it both} $(\xi,\eta_+)$
and $(-\xi,-\eta_+)$ lead to photon decay thresholds below 50 TeV.
The allowed region is the intersection of that from the old photon
decay constraint~\cite{Jacobson:2002hd,KonMaj} with its reflection
about the $\xi$ and $\eta$ axes. It is a pair of wedges in the
upper-right and bottom left quadrants. Numerical work shows that
this wedge pattern is maintained for different choices of $\eta_-$
relative to $\eta_+$, however the exact orientation and shape of
the wedges varies. A complete analysis of constraints would also
require examination of above threshold processes when the outgoing
particles have orbital angular momentum and hence helicities that
are not determined solely by the incoming photon.

%\paragraph
\subsubsection{Photon absorption}
\index{Lorentz violation constraints!QED at $O(E/M)$!photon absorption}
\index{photon absorption} A process related to photon decay is
photon absorption, $\g\g\rightarrow e^+e^-$. Unlike photon decay,
this is allowed in Lorentz invariant QED. If one of the photons
has energy $\omega_0$, the threshold for the reaction occurs in a
head-on collision with the second photon having the momentum
(equivalently energy) $k_{\rm LI}=m^2/\omega_{0}$. For $k_{\rm
LI}= 10$ TeV (which will be relevant for the observational
constraints) the soft photon threshold $\omega_0$ is approximately
25 meV, corresponding to a wavelength of 50 microns.

In the presence of Lorentz violating dispersion relations the
threshold for this process is in general altered, and the process
can even be forbidden. Moreover, as noticed by
Klu\'zniak~\cite{Kluzniak}, in some cases there is an upper
threshold beyond which the process does not occur.\footnote{
%-------------------------------------------------------
Our results agree with those of \cite{Kluzniak} only in certain
limiting cases.}
%-------------------------------------------------------
The lower and upper thresholds for photon annihilation as a
function of the two parameters $\xi$ and $\eta$ were obtained
in~\cite{Jacobson:2002hd}, before the helicity dependence required
by EFT was appreciated. As the soft photon energy is low enough
that its LV can be ignored, this corresponds to the case where
electrons and positrons have the same LV terms. The  analysis is
rather complicated. In particular it is necessary to sort out
whether the thresholds are lower or upper ones, and whether they
occur with the same or different pair momenta.

The photon absorption constraint, neglecting helicity dependent
effects, came from the fact that LV can shift the standard QED
threshold for annihilation of multi-TeV $\g$-rays from nearby
blazars, such as Mkn 501, with the ambient infrared extragalactic
photons~\cite{Kluzniak,GAC-Pir,SG,Jacobson:2002hd,KonMaj,Jacobson:2003ty,S03}.
LV depresses the rate of absorption of one photon helicity, and
increases it for the other.  Although the polarization of the
$\g$-rays is not measured, the possibility that one of the
polarizations is essentially unabsorbed appears to be ruled out by
the observations which show the predicted attenuation~\cite{S03}.
The electron and positron spin angular momenta add to at most one.
At threshold, where the collision is head-on, the photons must
therefore have opposite helicity, and hence the electron and
positron have opposite helicity. According to (\ref{eq:ep}), they
therefore have opposite LV parameters. The threshold analysis has
not been redone to account for this.

%\paragraph
\subsubsection{Vacuum photon splitting}
\index{Lorentz violation constraints!QED at $O(E/M)$!vacuum photon splitting}
\index{vacuum photon splitting} Another forbidden QED process that
is allowed in the presence of LV is vacuum photon splitting into
$N$ photons, $\g\rightarrow N\g$. Unlike the other processes
considered here, this would be a loop effect. The lowest order
Feynman diagram contributing would be a fermion loop with various
photon lines attached. The process has no threshold, so whether or
not it can be used to set constraints depends on the rate.

Aspects of vacuum photon splitting have been examined
in~\cite{Jacobson:2002hd, othersplitting}. An estimate of the
rate, independent of the particular form of the Lorentz violating
theory, was given in Ref.~\cite{Jacobson:2002hd}. It was argued
that a lower bound on the lifetime is $\d^{-4}E^{-1}$, where $\d$
is a Lorentz violating factor. For a photon of energy 50 TeV, this
is $10^{-29}\d^{-4}$ seconds. Such 50 TeV photons arrive from the
Crab nebula, about $10^{13}$ seconds away, so the best constraint
(i.e. if there is is no further small parameter such as $\a^N$ or
$1/16\pi^2$ in the decay rate) we could possibly get on $\d$ from
photon splitting is $\d\lesssim 10^{-10}$.

For a $p^n$ LV term with $n=2$ in the dispersion relation, this is
not competitive with the other constraints already obtained. For
higher $n$, each contribution arising from an operator of
dimension greater than four will be suppressed by at least one
inverse power of the scale $M$. For example, contributions from
$n=3$ would yield $\d\sim\xi E/M$. In this case the strongest
conceivable constraint on $\xi$ would be of order $\xi\lesssim
10^4$, and even this is not competitive with the other
constraints. \index{Lorentz violation constraints!QED at $O(E/M)$|)}

\subsection{Constraints at $O(E/M)$ from UHE cosmic rays}
\index{Lorentz violation constraints!UHE protons at $O(E/M)$}
\index{vacuum \v{C}erenkov effect} If ultra-high energy cosmic
rays (UHECR) are (as commonly assumed) protons, then we can derive
strong constraints on $n=3$ type dispersion by a) the absence of a
vacuum \v{C}erenkov effect at GZK energies and b) the position of
the GZK cutoff. For a soft emitted photon with a long wavelength,
the partonic structure of a UHECR proton is presumably irrelevant.
In this case we can treat the proton as a point particle as in the
QED analysis. With a GZK proton of energy $5 \times 10^{19}$ GeV
the constraint from the absence of a vacuum \v{C}erenkov effect is
$\eta< O(10^{-14})$.  For a hard emitted photon, the partonic
nature of the proton is important and the relevant mass scale will
involve the quark mass. The exact calculation considering the
partonic structure for $n=3$ has not been performed, however the
threshold region will be similar to that
in~\cite{Jacobson:2002hd}.  The allowed region in the $\eta-\xi$
plane will be bounded on the right by the $\xi$ axis (within a few
orders of magnitude of $10^{-14}$) and below by the line
$\xi=\eta$~\cite{Jacobson:2002hd}. These constraints apply to only
one helicity of proton and photon, since the UHECR could consist
all of a single helicity. Also the different quarks could have
different dispersion parameters. See however section~\ref{M2} for
remarks on the approach of~\cite{Gagnon:2004xh} which can be
applied to deduce combined constraints in this case.

\index{GZK cutoff!and Lorentz violation}If the GZK cutoff is
observed in its predicted place, this will place limits on the
parameters $\eta_p$ and $\eta_\pi$. For example, if the GZK cutoff
is eventually observed to be somewhere between 2 and 7 times
$10^{19}$ GeV then there are strong constraints of $O(10^{-11})$
on $\eta_p$ and $\eta_\pi$~\cite{Jacobson:2002hd}.  As a final
comment, an interesting possible consequence of LV is that with
upper thresholds, one could possibly reconcile the AGASA and
Hi-Res/Fly's Eye experiments.  Namely, one can place an upper
threshold below $10^{21}$ GeV while keeping the GZK threshold
near $5\times 10^{19}$ GeV.  Then the cutoff would be
``seen'' at lower energies but extra flux would still be present at
energies above $10^{20}$ GeV, potentially explaining the AGASA
results~\cite{Jacobson:2002hd}.  The region of parameter space for
this scenario is terribly small, however, again of $O(10^{-11})$.

\subsection{Constraints at $O(E^2/M^2)$?}
\label{M2}
\index{Lorentz violation constraints!$O(E^2/M^2$)|(}
As previously mentioned, CPT
symmetry alone could exclude the dimension five LV operators that
give $O(E/M)$ modifications to particle dispersion relation, and
in any case the constraints on those have become nearly
definitive. Hence it is of interest to ask about the $O(E^2/M^2)$
modifications. We close with a brief discussion of the constraints
that might be possible on those, i.e. constraints at $O(E^2/M^2)$.

\index{Lorentz violation constraints!$O(E^2/M^2$)!QED}
As discussed above, the
strength of constraints can be estimated by the requirement
$\eta_4 p^4/M^2 \lesssim m^2$, which yields
\beq \eta_4\lesssim \left(\sqrt{\frac{m}{1\, {\rm eV}}}\frac{100\,
{\rm TeV}}{p}\right)^4. \label{eta4bound} \eeq
Thus, for electrons, an energy around $10^{17}$ eV is needed for
an order unity constraint on $\eta_4$, and we are probably not
going to see any effects directly from such electrons.

\index{Lorentz violation constraints!$O(E^2/M^2$)!UHE cosmic rays}
For protons an
energy $\sim 10^{18}$ eV is needed. This is well below the UHE
cosmic ray energy cutoff, hence if and when Auger~\cite{Auger}
confirms the identity of UHE cosmic rays as protons at the GZK
cutoff, we will obtain an impressive constraint of order
$\eta_4\lesssim 10^{-5}$ from the absence of vacuum \v{C}erenkov
radiation for $10^{20}$ eV protons.  From the fact that the GZK
threshold is not shifted, we will obtain a constraint of order
$\eta_4\gtrsim -10^{-2}$, assuming equal $\eta_4$ values for
proton and pion.

In fact, if one assumes the cosmic rays near but below the GZK
cutoff are hadrons, one already obtains a strong
bound~\cite{Gagnon:2004xh}. Depending on the species dependence of
the LV coefficients, bounds of order $10^{-2}$ or better can be
placed on $\eta_4$.  The bounds claimed in ~\cite{Gagnon:2004xh}
are actually two sided, and it is worthwhile to explain how such
bounds come about for a single source particle.  Up to this point
it has been necessary to use at least \textit{two} reactions with
different source particles to derive a two sided bounds.  For
example, the Crab constraints rely on the existence of both 50 TeV
electrons and photons, treating each as a fundamental particle
with its own LV coefficient.  In contrast, the two sided bounds
in~\cite{Gagnon:2004xh} are derived by using a parton model for
particles where the LV coefficients apply to the constituent
partons.  By considering many different outgoing particle spectra
from an incoming hadron in combination with the parton approach
the authors of~\cite{Gagnon:2004xh} are able to find sets of
reactions that yield two sided bounds.  Hence, the parton approach
is extremely useful as it dramatically increases the number of
constraints that can be derived from a single incoming particle.
However, it also requires more assumptions about the behavior of
the parton distributions at cosmic ray energies.

\index{Lorentz violation constraints!$O(E^2/M^2$)!UHE neutrinos}
\index{neutrino \v{C}erenkov radiation}
Impressive constraints might also be obtained from the absence of neutrino
vacuum \v{C}erenkov radiation: putting in 1 eV for the mass in
(\ref{eta4bound}) yields an order unity constraint from 100 TeV
neutrinos, but only if the \v{C}erenkov {\it rate} is high enough.
The rate will be low, since it proceeds only via the non-local
charge structure of the neutrino. Recent calculations~\cite{Dave?}
have shown that the rate is not high enough at that energy.
\index{UHE neutrinos}
However, for $10^{20}$ eV UHE neutrinos, which may be observed by
the proposed EUSO and/or OWL satellite observatories, the
rate will be high enough to derive a strong constraint.  The
value of the constraint would
depend on the emission rate, which has not yet been
computed.  For a {\it gravitational} \v{C}erenkov reaction, the
rate (which is lower but easier to compute than the
electromagnetic rate) would be high enough for a neutrino from a
distant source to radiate provided $\eta_4\gtrsim 10^{-2}$. Hence in this
case one might obtain a constraint of order $\eta_4\lesssim
10^{-2}$, or stronger in the electromagnetic case.

\index{Lorentz violation constraints!$O(E^2/M^2$)!time of flight}
A time of flight constraint at order $(E/M)^2$ might be
possible~\cite{GACsecgen} if gamma ray bursts produce UHE ($\sim
10^{19}$ eV) neutrinos, as some models predict, via limits on time
of arrival differences of such UHE neutrinos vs. soft photons (or
gravitational waves). Another possibility is to obtain a vacuum
birefringence constraint with higher energy photons~\cite{Mitro},
although such a constraint would be less powerful since EFT does
not imply that the parameters for opposite polarizations are
opposite at order $(E/M)^2$. If future GRB's are found to be
polarized at $\sim 100$ MeV, that could provide a birefringence
constraint $|\xi_{4+}-\xi_{4-}|\lesssim 1$.

\index{Lorentz violation constraints!$O(E^2/M^2$)|)}

\index{Lorentz violation constraints|)}

\section{Conclusion}

At present there are only hints, but no compelling evidence for
Lorentz violation from quantum gravity. Moreover, even if LV is
present, the use of EFT for its low energy parametrization is not
necessarily valid. Nevertheless, we believe that
the constraints derived from the simple ideas discussed
here are still important.  They allow tremendous advances in
observational reach to be applied in a straightforward manner to
limit reasonable possibilities that might arise from fundamental
Planck scale physics. Such guidance is especially welcome for the
field of quantum gravity, which until the past few years has had
little connection with observed phenomena.

%----------------------------

%
\printindex
\end{document}